\documentclass[screen,nonacm,authorversion]{acmart}

\AtBeginDocument{%
  }

\setcopyright{none}
\copyrightyear{2025}
\acmYear{2025}
\acmDOI{XXXXXXX.XXXXXXX}

\acmConference[CGO '25]{Make sure to enter the correct
  conference title from your rights confirmation email}{
  February, 2025}{Las Vegas, NV}

\acmISBN{978-1-4503-XXXX-X/18/06}

\usepackage{etoolbox}
\usepackage{hyperref}
\usepackage{cleveref}
\usepackage[lining]{FiraMono}
\usepackage{amsmath}

\usepackage{circledsteps} 

\usepackage{subcaption}

\usepackage{listings}
\lstdefinestyle{MLIR}{%
  keywords = {
    arith.constant,
    func.func,
    memref.store,
    memref.subview,
    scf.forall,
  },
}
\lstdefinestyle{IRDL}{%
  keywords = {
    Attributes,
    Dialect,
    CPPConstraint,
    Operation,
    Operands,
    Results,
    Regions,
  },
}
\lstdefinestyle{transform}{%
  keywords=[1]{
    transform.alternatives,
    transform.apply_patterns,
    transform.pattern.add_of_zero_pad,
    transform.pattern.negate_of_transpose,
    transform.pattern.matmul_of_transpose,
    transform.named_sequence,
    transform.expand_strided_metadata,
    transform.to_library,
    transform.vectorize,
    transform.assert,
    transform.to_library
    match.op,
    loop.split,
    loop.tile,
    loop.unroll,
  },
  keywords=[2]{
    tile0, tile1, tile2, tile3, vect,
  },
  keywordstyle=[2]\itshape\color{purple},
}
\lstdefinestyle{OpenMP}{%
  keywords=[1]{
    pragma,
    omp,
    tile,
    sizes,
    simd,
  },
  keywords=[2]{
    tile0, tile1, tile2, tile3, vect,
  },
  keywordstyle=[2]\itshape\color{purple},
}

\usepackage{tikz}

\usetikzlibrary{tikzmark}
\usetikzlibrary{calc}
\usetikzlibrary{positioning}
\usetikzlibrary{arrows}
\usepackage{pgfplots}
\usetikzlibrary{pgfplots.groupplots}

\newtoggle{draftComments}
\togglefalse{draftComments}

\iftoggle{draftComments}{%
  \addtolength{\marginparwidth}{1.4in}
  \addtolength{\marginparsep}{0.3in}
  \addtolength{\hoffset}{1.5in}
  \addtolength{\paperwidth}{3.0in}
  \newcommand{\extras}[1]{{\color{blue}{#1}}}%
  \newcommand{\margincomment}[1]{#1}
}{%
  \newcommand{\extras}[1]{}%
  \newcommand{\margincomment}[1]{}
}

\newcommand{\az}[1]{\margincomment{\marginpar{\color{brown}\small\textbf{Alex:} #1}}}

\newcommand{\ms}[1]{\margincomment{\marginpar{\color{orange}\small\textbf{Michel:} #1}}}

\usepackage{letltxmacro}

\LetLtxMacro\oldttfamily\ttfamily
\DeclareRobustCommand{\ttfamily}{\oldttfamily\csname ttsize\endcsname}
\newcommand{\setttsize}[1]{\def\ttsize{#1}}%
\setttsize{\footnotesize}

\newcommand{\TODO}[1]{{\color{red}\textbf{TODO:} #1}}

\title{The MLIR Transform Dialect}

\subtitle{Your compiler is more powerful than you think}

\author{Martin Paul Lücke}
\email{martin.luecke@ed.ac.uk}
\affiliation{
  \institution{University of Edinburgh}
  \city{Edinburgh}
  \state{Scotland}
  \country{UK}
}

\author{Oleksandr Zinenko}
\email{contact@ozinenko.com}
\affiliation{
  \institution{Google DeepMind}
  \city{Paris}
  \country{France}
}

\author{William S. Moses}
\email{wsmoses@illinois.edu}
\affiliation{
  \institution{Google DeepMind}
  \city{Cambridge}
  \state{MA}
  \country{~}
}
\affiliation{
  \institution{University of Illinois Urbana-Champaign}
  \state{IL}
  \country{United States}
}

\author{Michel Steuwer}
\email{michel.steuwer@tu-berlin.de}
\affiliation{
  \institution{TU Berlin}
  \city{Berlin}
  \country{Germany}
}

\author{Albert Cohen}
\email{albertcohen@google.com}
\affiliation{
  \institution{Google DeepMind}
  \city{Paris}
  \country{France}
}

\begin{document}

\begin{abstract}

To take full advantage of a specific hardware target, performance engineers need to gain control on compilers in order to leverage their domain knowledge about the program and hardware.
Yet, modern compilers are poorly controlled, usually by configuring a sequence of coarse-grained monolithic black-box passes, or by means of predefined compiler annotations/pragmas. These can be effective, but often do not let users precisely optimize their varying compute loads.
As a consequence, performance engineers have to resort to implementing custom passes for a specific optimization heuristic, requiring compiler engineering expert knowledge.

In this paper, we present a technique that provides fine-grained control of general-purpose compilers by introducing the \emph{Transform dialect}, a controllable IR-based transformation system implemented in MLIR.
The Transform dialect empowers performance engineers to optimize their various compute loads by composing and reusing existing---but currently hidden---compiler features without the need to implement new passes or even rebuilding the compiler.

We demonstrate in five case studies that the Transform dialect enables precise, safe composition of compiler transformations and allows for straightforward integration with state-of-the-art search methods.

\end{abstract}

\maketitle

\section{Introduction}

Compilers are typically assembled as a sequence of passes, or \emph{pass pipeline}, augmented with flags to configure this pipeline and influence heuristics.
These passes are subsequently run on intermediate representations (IR) of a program until the compilation process concludes.
This approach to controlling compilers enables users to order the passes of the compiler, and to perform specific optimizations parameterized by their corresponding flags--- e.g.\ apply \emph{loop invariant code motion} on all loops.

However, this coarse level of control is increasingly insufficient to optimize programs for today's heterogeneous hardware that require precise optimization decisions.
Pragmas, or compiler annotations in the source code, provide finer grained control---e.g.\ vectorization or unrolling hints. These are effective but their implementation requires in-depth and non-modular changes to the compiler, hence their restriction to specific cases anticipated by compiler engineers.

Often specific parts of a program dominate the overall runtime and are worth optimizing precisely or offloading to an accelerator. In this case, the user, an expert in their application domain, requires the ability to communicate their knowledge on how the program is to be optimized to the compiler.
Traditionally, this is performed manually on the program itself by intertwining the algorithm and its optimization using a low-level programming language like C.
This style severely limits the portability of a program, often requiring writing and optimizing it again for different target hardware.

Additionally, many specialized domains with complex software stacks such as machine learning only offer high-level programming models that do not allow expressing optimizations on the required lower level of abstraction.

Recent domain-specific scheduling approaches such as Halide~\cite{halide} and TVM~\cite{tvm} address these concerns in the specific domains of image processing and machine learning. They enable expressing a \emph{schedule} separately from the program to specify its optimization. As a result, supporting a new hardware target only requires a new schedule, not a complete redesign of the program.
This separation of concerns facilitates reuse and enables the integration of automatic parameter search methods and schedule synthesis~\cite{zheng2020ansor}.
\az{We need a proper citation for synthesis here, not sure if Ansor is the right one.}

These existing domain-specific approaches introduce their own software stacks specific to their domain and do not integrate well with existing compiler infrastructures. For instance, reusing an existing compiler optimization requires implementing it again in the respective framework.

While modern compiler construction infrastructure such as MLIR contain similar transformations as implemented in the Halide and TVM compilers, these do not get exposed to users. For instance, MLIR contains functions that implement tiling, loop interchange or loop unrolling but only exposes them bundled as a pass which has to be applied to all loops. These passes typically follow a specific heuristic that yields good results on certain benchmarks but leaves users without further control.
Realizing a specific composition of transformations, akin to a schedule, using the existing functions in MLIR requires users to write new passes in the compiler source language and rebuild the compiler. A process that requires expert knowledge beyond the domain of the program the user aims to express and optimize.

With \emph{Transform dialect}, we present an approach to exposing existing, but currently hidden compiler features and precisely composing them to control the optimization of programs.

We represent our transformation language for controlling compiler optimizations directly as compiler IR.

With our approach the input to compilers consists of two parts: (1) the computational IR 
describing the computation being optimized, referred to as the \emph{payload program}, and (2) the Transform script for controlling the compiler. The compilation process is driven by interpreting the Transform IR that describes how to gradually transform and optimize the payload program.

The Transform dialect is implemented in MLIR and contains a base set of operations to model compositions of transformations. It follows an extensible design to expose new transformations.

We extend MLIR with an interface to make existing helper functions used in passes accessible to Transform scripts.
Thus, exposing fine-grained transformation steps without compromising on their usability from other native code.

By introducing the Transform dialect, we open up the hidden features of general-purpose compilers to non-compiler experts in order to enable precisely injecting domain knowledge into the compilation process.

Our contributions are:
\begin{itemize}
    \item The Transform dialect, an extensible approach for fine-grained control of compiler transformations. (\Cref{sec:transform_dialect})
    \item A system of pre- and post-conditions to statically detect problems in compiler pipelines (\Cref{sec:pre-post-conditions})
    \item An evaluation of the Transform dialect with five case studies (\Cref{sec:case_studies}) highlighting its low computational overhead and robustness, as well as its usefulness for detecting performance problems, generating high-performance code and exploring optimization spaces.
\end{itemize}

\section{Motivation and Background}

In the past, compilers were mostly designed to optimize for common cases.
This made sense in a world where applications were written in one of a few general-purpose programming languages, such as C, C++ and Fortran.
Compiler engineers could also focus on a relatively limited class of heuristics, as code was generated only for one of few popular computer architectures, such as x86 and ARM.

This picture has already become one of the distant past.
Nowadays, software is increasingly written in domain-specific languages and software stacks. This has sparked the development of domain-specific compilers, such as Halide~\cite{halide}, TVM~\cite{tvm}, TACO~\cite{taco}, Lift~\cite{lift} \& RISE~\cite{rise}, and more, but also of new compiler frameworks to facilitate their development, most prominently MLIR~\cite{lattner2020mlir}.
Our hardware landscape has evolved as well, resulting in a growing diversity of accelerator architectures including GPUs, Google's TPU, Groq's LPU, or Cerebras' wafer-scale engine, only to name a few.

Optimizing programs in this new landscape is more challenging as it requires tweaking and adjusting optimizations for the target hardware architecture as well as the specific computation performed by the input program.
But, we know from the domain of machine-learning, that performing optimizations specific to individual computational kernels can lead to significant performance gains that can not be ignored.

\begin{figure*}[t]
\begin{subfigure}{0.98\textwidth}
\begin{lstlisting}[basicstyle=\fontseries{uluc}\fontsize{5.5}{7.5}\selectfont\ttfamily, keywordstyle = \fontseries{uluc}\fontsize{5.5}{7.5}\selectfont\ttfamily\bfseries, numbers=left,numbersep=5pt,xleftmargin=.4cm]
named_sequence @split_then_tile_and_unroll(%func) {
  |\tikzmark{outer}|%outer|\tikzmark{outer_end}|     = match.op "scf.for" {first} in %func         // type: outer:    {scf.for}
  |\tikzmark{hoisted}|%hoisted|\tikzmark{hoisted_end}|   = loop.hoist from %outer to %func             // type: hoisted:  {*.*}
  |\tikzmark{inner}|%inner|\tikzmark{inner_end}|     = match.op "scf.for" {first} in %outer        // type: inner:    {scf.for}
  %param     = param.constant 8
  %part:2    = loop.split %inner ub_div_by=%param          // type: part#1:   {scf.for[ub%8=0]}
                                                           //       part#2:   {scf.for[ub<8]}
  |\tikzmark{tiled}|%tiled:2|\tikzmark{tiled_end}|   = loop.tile %part#1 tile_sizes=[%param].      // type: tiled#1:  {scf.for[part#1.ub/8]}
                                                           //       tiled#2:  {scf.for[ub=8]}
  |\tikzmark{unroll}|%unrolled|\tikzmark{unroll_end}|  = loop.unroll %part#2 {full}                  // type: unrolled: {*.*}
  %unrolled2 = loop.unroll %part#2 {full}                  // This statically reports an error!
}

\end{lstlisting}

\caption{Operations of the Transform dialect are used to express a tiling optimization of an uneven inner loop. Inputs and outputs to transforms are represented explicitly to enable precise chaining. Metadata such as tile sizes are used to further configure individual transforms. Static reasoning about the possible structure of the payload IR is shown in comments on the right hand side.}
\label{fig:transformIR}
\end{subfigure}
\par\bigskip
\begin{subfigure}{0.49\textwidth}
\begin{lstlisting}[basicstyle=\fontseries{uluc}\fontsize{5.5}{7.5}\selectfont\ttfamily, keywordstyle = \fontseries{uluc}\fontsize{5.5}{7.5}\selectfont\ttfamily\bfseries, numbers=left,numbersep=5pt,xleftmargin=.4cm]
func @myFunc(%values: memref) {


  |\tikzmark{outer_p1}|scf.for %i = 0 to 4096|\tikzmark{outer_p1_end}| {
    |\tikzmark{hoisted_p1}|%c1 = arith.constant 1|\tikzmark{hoisted_p1_end}|
    |\tikzmark{inner_p1}|scf.for %j = 0 to 2042|\tikzmark{inner_p1_end}| {
      |\tikzmark{hoisted_p1_2}|%c2 = arith.constant 2|\tikzmark{hoisted_p1_2_end}|
    
      %val = memref.load %values[%c1, %i, %j]
      func.call @use(%val, %c2)
    }
    
    
    
    
    
  }  }
\end{lstlisting}
\caption{Initial payload IR\\[-1em]}
\label{fig:payloadIR_before}
\end{subfigure}
\begin{subfigure}{0.49\textwidth}
\begin{lstlisting}[basicstyle=\fontseries{uluc}\fontsize{5.5}{7.5}\selectfont\ttfamily, keywordstyle = \fontseries{uluc}\fontsize{5.5}{7.5}\selectfont\ttfamily\bfseries, numbers=left,numbersep=5pt,xleftmargin=.4cm]
func @myFunc(%values: memref) {
  |\tikzmark{hoisted_p2}|%c1 = arith.constant 1|\tikzmark{hoisted_p2_end}|
  |\tikzmark{hoisted_p2_2}|%c2 = arith.constant 2|\tikzmark{hoisted_p2_2_end}|
  |\tikzmark{outer_p2}|scf.for %j = 0 to 4096|\tikzmark{outer_p2_end}| {
  
    |\tikzmark{tiled_1}|scf.for %i_0 to 255|\tikzmark{tiled_1_end}| {
      |\tikzmark{tiled_2}|scf.for %i_1 = 0 to 8|\tikzmark{tiled_2_end}| {
        %i   = arith.muli %i_0, %i_1
        %val = memref.load %values[%c1, %i, %j]
        func.call @use(%val, %c2)
      }
    }
    |\tikzmark{unroll_1}|%val2020 = memref.load %values[%c, 2040, %j]|\tikzmark{unroll_1_end}|
    |\tikzmark{unroll_2}|func.call @use(%val)|\tikzmark{unroll_2_end}|
    |\tikzmark{unroll_3}|%val2021 = memref.load %values[%c, 2041, %j]|\tikzmark{unroll_3_end}|
    |\tikzmark{unroll_4}|func.call @use(%val)|\tikzmark{unroll_4_end}|
  }  }
\end{lstlisting}
\caption{Transformed payload IR\\[-1em]}
\label{fig:payloadIR_after}
\end{subfigure}
\caption{A Transform script performing code hoisting, loop splitting, tiling, and unrolling is defined (a) and used to transform the initial payload IR (b) to the transformed IR (c). The operations associated with specific handles are colored similarly.}
\label{fig:transformIR_payloadIR}

\begin{tikzpicture}[remember picture,overlay, transform canvas={yshift=0.25em}]
\begin{scope}[line width=7.5pt, color=green, opacity=0.3]
    \draw (pic cs:outer) -- (pic cs:outer_end);
    \draw (pic cs:outer_p1) -- (pic cs:outer_p1_end);
    \draw (pic cs:outer_p2) -- (pic cs:outer_p2_end);
\end{scope}
\begin{scope}[line width=7.5pt, color=brown, opacity=0.3]
    \draw (pic cs:hoisted) -- (pic cs:hoisted_end);
    \draw (pic cs:hoisted_p1) -- (pic cs:hoisted_p1_end);
    \draw (pic cs:hoisted_p1_2) -- (pic cs:hoisted_p1_2_end);
    \draw (pic cs:hoisted_p2) -- (pic cs:hoisted_p2_end);
    \draw (pic cs:hoisted_p2_2) -- (pic cs:hoisted_p2_2_end);
\end{scope}
\begin{scope}[line width=7.5pt, color=yellow, opacity=0.3]
    \draw (pic cs:inner) -- (pic cs:inner_end);
    \draw (pic cs:inner_p1) -- (pic cs:inner_p1_end);
\end{scope}
\begin{scope}[line width=7.5pt, color=lime, opacity=0.3]
    \draw (pic cs:tiled) -- (pic cs:tiled_end);
    \draw (pic cs:tiled_1) -- (pic cs:tiled_1_end);
    \draw (pic cs:tiled_2) -- (pic cs:tiled_2_end);
\end{scope}
\begin{scope}[line width=7.5pt, color=olive, opacity=0.3]
    \draw (pic cs:unroll) -- (pic cs:unroll_end);
    \draw (pic cs:unroll_1) -- (pic cs:unroll_1_end);
    \draw (pic cs:unroll_2) -- (pic cs:unroll_2_end);
    \draw (pic cs:unroll_3) -- (pic cs:unroll_3_end);
    \draw (pic cs:unroll_4) -- (pic cs:unroll_4_end);    
\end{scope}
\end{tikzpicture}
\end{figure*}

\subsection{Controlling Transformations Today in MLIR}
MLIR~\cite{lattner2020mlir} has emerged as a popular framework for building domain-specific compilers.
It closely follows LLVM~\cite{llvm}, but crucially allows to define custom abstractions for representing specific computations as well as custom compiler transformations.
Programs are represented in a hierarchical SSA-based form\footnote{\url{https://mlir.llvm.org/docs/LangRef}} and are structured as control-flow graphs (CFG) of control-flow-free basic \emph{blocks}.
Each block has a set of \emph{block arguments} and contains a sequence of \emph{operations}, each of which can produce multiple return \emph{values}.
Block arguments and values representing the results of prior operations are used as the operation's \emph{operands}.
Finally, operations might have \emph{regions} containing a hierarchically nested CFG.

MLIR is extremely customizable, allowing users to define their own operations and types, arranged into libraries called \emph{dialects}. It provides a rich plugin mechanism allowing new definitions to be injected without recompiling the compiler, as dynamic libraries or using declarative specifications~\cite{irdl}.

Compiler transformations in MLIR are usually bundled into \emph{passes}, following the design of LLVM.
Passes in modern compilers are designed to be reusable in different pass pipeline configurations, enabling performance engineers to explore which pass pipeline configuration leads to the best performance for their program and target scenario.

However, the control via passes is coarse-grained as there is no universal way to restrict passes to a targeted part of the program.
Furthermore, rearranging pass pipelines is restricted by implicit dependencies between passes that are not explicitly expressed beyond textual documentation.

\label{sec:linalg}
A significant driver of the implementation of individual optimizations in MLIR is the domain of linear algebra computations found in ML workloads, particularly those expressed using \emph{structured operations} in the Linalg and related dialects~\cite{linalg}. Initially, high-impact transformations such as tiling and fusion, were orchestrated using a system designed for peephole optimization~\cite{peephole} of static single assignment IRs, borrowing ideas from term rewriting.
\az{Citations for term rewriting, LLVM if there's something relevant}
In this style, each transformation is expressed as a \emph{rewrite pattern}, implemented in an imperative style using C++ code, that matches relevant operations and replaced them with a transformed equivalent.
As more fine-grained controls are lacking, such patterns are applied greedily to the entire program until a fixed point is reached. 
This approach has led to:
\emph{1)} rewrite patterns that progressively include target-specific hard-coded heuristics, e.g., only to unroll loops with less than 8 iterations;
\emph{2)} conflicts between different rewrite patterns that are applicable to the same inputs, resulting in situations where it is unclear, e.g., whether a loop should be first tiled or fused with another loop.

As a remedy, such patterns started relying on IR metadata to externalize heuristic decisions (e.g., this loop should be tiled with size 32), establish order (e.g., this operation has been tiled and should not be tiled again) or otherwise communicate with each other. However, this approach is brittle as metadata is not guaranteed to be preserved by transformations.
Furthermore, metadata may not reference IR constructs to which it is not directly attached, requiring delicate string-based matching and additional IR traversals to link together several IR pieces.
\az{Consider referencing specific commits or RFCs.}

We need a more principled approach to facilitate the fine-grained control of compiler optimizations in MLIR.

\subsection{Scheduling Languages for Separating Computations and Optimization Decisions}

Halide~\cite{halide} pioneered the idea of separating computation from a \emph{schedule} that specifies the sequence of optimizations to be performed.
This approach has the advantage of portability: while the computation remains unchanged, different schedules describe different optimizations for different hardware targets, or even inputs.
This model has since been popularised by other domain-specific compilers, including TVM and TACO.
Unfortunately, all these solutions are domain-specific and not accessible to a generic compiler framework.

ELEVATE~\cite{rise} demonstrates how to build a generic scheduling language from first principles using formal programming language foundations.
However, this formal rewrite-based approach requires the computational program to be side-effect free --- a significant restriction that is unrealistic to assume or to ensure in many practical settings.

\smallskip
Therefore, we see a need for a scheduling language for a generic compiler framework to enable the fine-grained control of compiler optimizations for a wide variety of settings.
In the following, we describe such a scheduling language for MLIR implemented as the \emph{Transform dialect}.

\section{Transform Dialect: Representing and Controlling Transformations using IR}
\label{sec:transform_dialect}

We introduce the Transform dialect by example, designing an optimization as composition of existing transformations and applying it to a program as shown in \Cref{fig:transformIR_payloadIR}.
The Transform script shown in \Cref{fig:transformIR} first hoists code out of a specific loop (line~3), before splitting a nested loop into two loops (line~6), then tiles the first resulting loop (line~8) and unrolls the second (line~10).
The script contains a deliberate error (line~11) where we attempt to unroll the same loop a second time, to highlight that such errors are detected statically.

This script is executed sequentially from top to bottom and consists of MLIR operations that we call \emph{transforms}. A transform may directly model a transformation, such as in lines 3, 6, 8, 10, 11 where part of the payload program is rewritten, or a transform may assist in targeting or composing other transforms, for instance by matching a nested operation such as in lines 2 and 4. Transforms define values called \emph{handles}, each of which refers to a list of operations in the payload IR.
As Transform scripts are represented in MLIR itself, handles are regular MLIR values that obey the usual Static Single Assignment (SSA) rules. Other transforms may use handles as operands in order to transform the associated payload operations or extract information from them. The relation of transforms and their associated payload operations is indicated in the example through similar coloring. Additional operations with regions may be used to organize the Transform script into conditionals, loops or even functions.

In addition to handles, the Transform script may use \emph{parameters}. These values may be unknown when the Transform script is created but known when it is executed. Since they also obey SSA, parameters are effectively constant during execution. Together with attributes on the transform operations, parameters specialize the operation by providing, e.g., sizes for loop tiling or the preferred vector width. \Cref{fig:transformIR} uses a constant parameter in line~5 to specify split and tile sizes. This approach makes the Transform dialect a medium for \emph{externalizing compiler heuristics}: instead of parameterizing a transform in the compiler code, one can now generate Transform scripts with function-like abstractions that accept parameters as operands and use them inside the transformation. Parameters can also be derived from the payload IR. For example, an operation accepting a loop handle and producing a parameter with desired tile sizes is perfectly suitable for the Transform dialect.

Transform scripts are executed by the compiler when compiling the program via an interpreter that maintains the association table between handles and payload operations and dispatches execution to transformation logic implemented in C++ using MLIR interfaces.\footnote{MLIR interfaces provide dynamic polymorphism, similarly to OOP.} While our current interface-based approach is designed for compiler extensibility, Transform scripts can also be lowered down to LLVM IR, (JIT\nobreakdash-) compiled and linked to compiler libraries should compiler performance become a critical concern.

Since transformations may not always succeed, the transform interpreter provides an error handling mechanism similar to exceptions.\footnote{As of the time of writing, MLIR does not model exceptions. We resort to implicit error-checking semantics in all transforms. All transforms check if an error had been signaled and are implicitly no-ops in that case.} A transform may signal a silenceable or a definite error. In the former case, the interpreter will skip the remainder of the current region and return control flow to the parent transform operation. This operation may decide to suppress the error or report it for further handling. Silenceable errors typically indicate a failed precondition or at least that the payload has not been modified irreversibly. Definite errors cannot be suppressed and are immediately reported, aborting the interpreter.

A detailed description of the Transform dialect extensibility mechanisms is available in the documentation.\footnote{\url{https://mlir.llvm.org/docs/Tutorials/transform/}}

\subsection{Dealing with a Mutable IR}
\label{sec:handle_invalidation}
Contrary to many purely functional approaches~\cite{lift,rise,egg}, the Transform dialect is embedded into a practical compiler infrastructure operating on a payload IR that is implemented mutably for efficiency reasons. A transformation in MLIR may erase the payload operation to which a handle is pointing, either to replace it with a newly created operation or to remove it irreversibly, leaving the handle \emph{dangling}.
To prevent invalid access through such handles, we introduce the notion of \emph{handle invalidation}, akin to the concept of iterator invalidation known in C++.
Therefore, a transform must indicate whether it invalidates its operands, which internally corresponds to indicating a ``memory deallocation'' side effect on the operation.
Invalidating a handle also invalidates any other handles to the same payload or any of its parts, e.g., a nested operation.
Invalidated handles cannot be used as operands anymore.

To enable chaining of transforms, transforms that invalidate their operands typically return new handles to their results, such as in lines 6, 8 in \Cref{fig:transformIR}.
For instance, a loop reversal transformation would invalidate the handle to the loop and return the handle to the reverted loop if its implementation recreates the loop operation, and could preserve the handle if the reversal is done in place.
Returning new handles from a transformation brings the representation conceptually close to functional approaches, especially given the single-assignment property of the IR~\cite{appel1998ssa}, but the underlying payload IR remains mutable.

To help the user avoid accessing invalidated handles in the Transform script, the interpreter keeps track of invalidation, including handles invalidated indirectly due to nesting in the payload IR discovered by traversing the payload IR along with the handle/operation mapping, and reports errors. Static analysis is also possible as described in \Cref{sec:transform_ir_benefits}.

Additionally, a transform may subscribe to ``operation replaced'' and ``erased'' events from the MLIR rewrite driver, used by large passes such as dialect conversions and peephole optimizations. In reaction to these events, the transform may prevent handle invalidation by updating the handle to point to the replacement operation or to point to the empty set of operations, depending on the desired logic.

\subsection{Extensibility}
\label{sec:extensibility}
\az{We may want an inverse take on this: how does one orchestrate transforms in the system as extensible as MLIR?}
The Transform dialect builds on MLIR's extensibility, allowing advanced users to define new transformation operations associated either with existing or new custom IR transformations implemented in the compiler.
These transforms can naturally mix with each other, as any other dialect.

When modifications of the compiler are undesirable, too challenging or simply impossible, one can nevertheless create new transform abstractions as combinations of existing transforms. Indeed, since transforms are mere operations, they can be organized into macros or functions using abstractions already present in MLIR. For example, the \texttt{named\_sequence} operation on line~1 of \Cref{fig:transformIR} defines a macro that can be expanded in any other place using the \texttt{include} operation.\footnote{In the implementation, macros are preferred to functions in many cases as they lead to simpler flow in the interpreter.}

In Halide and TVM, the schedule and computational program are written closely side-by-side sharing the same variable scope and the schedule can directly refer to computational variables.
But this means also that schedules are specific to a single program.
In contrast, in our approach Transform scripts are not program specific and separated from the computational program making the compositions of transforms easily reusable and further composable.
This in turn opens the door for building libraries of composed compiled transforms and distributing them, potentially separately from the compiler.

Furthermore, since transforms are represented as ordinary compiler IR, they can be subject to compiler rewrites themselves. For example, a loop tiling transformation itself can be ``lowered'' to the canonical combination of loop strip-mining and interchange instead of being implemented directly, and these loop interchanges may cancel out with eventual further interchanges or other transforms present using regular pattern-driven peephole optimization. 

\subsection{Composability with Pre- and Post-Conditions}
\label{sec:pre-post-conditions}
Most compiler transformations come with assumptions about the input IR that are required for the transformation to be successfully applied. When these pre-conditions are not satisfied, defensively-written transformations will not modify the IR and potentially warn the user, but poorly-written transformations may result in miscompilations or introduce subtle bugs. The Transform dialect provides a natural place to explicitly declare pre-conditions as well as specify post-conditions guaranteed by the transformation to ensure working compositions. Pre- and post-conditions are expressed using transform operation attributes as well as the types of handles, both of which are fully user-extensible in MLIR.

One of the most common uses for the pre- and post-conditions is for a set of \emph{progressive lowering} transformations. Each of these removes a subset of operations belonging to one dialect and introduces new operations, potentially from another ``lower'' dialect. For example, a \texttt{convert\_scf\_to\_cf} lowering transform replaces structured control flow (SCF) dialect operations, such as loops and conditionals, with classical branch-based control flow defined in the CF dialect. Building a full compilation flow in MLIR requires composing multiple lowerings until all operations use the desired set of the final target dialects, such as LLVM IR or SPIR-V dialects.
Determining the right order of lowerings is usually an expensive, trial-and-error process subject to the phase ordering issues~\cite{mlir-not-ml-compiler} as there are no programmatically-accessible information about the operations that are being lowered out and those that are being introduced.

The Transform dialect allows developers to make pre- and post-conditions explicitly available by listing operations that are being added and removed by each lowering transform as shown in \Cref{fig:scf_to_cf_constraints}. In this example, the next lowerings should take care of converting CF and arithmetic operations towards the desired target dialects. One can also statically observe that any loop transformations operating on SCF, such as interchange or unrolling, must be ordered before \texttt{convert\_scf\_to\_cf} allowing to check for phase ordering violations statically.
We have implemented a prototype tool for checking statically if a composition of transforms violates the specified pre-conditions of any individual transform.

\begin{figure}[t]
\centering
\begin{lstlisting}[escapeinside={!}{!},morekeywords={transform.convert_scf_to_cf, cf., scf, arith}]
transform.convert_scf_to_cf(input: {scf.!*!}) ->
  (result: {cf.branch, cf.switch, cf.cond_branch,
        arith.addi, arith.cmpi, arith.index_cast})
\end{lstlisting}
\caption{Pre-/post-conditions of \texttt{convert\_scf\_to\_cf} declare which kinds of payload operations are consumed and removed (all operations from the \texttt{scf} dialect), and which new operations are introduced by this transform (the operations listed explicitly in lines 2 and 3).}
\label{fig:scf_to_cf_constraints}
\end{figure}

To further facilitate extensibility, it is also possible to not list specific operation names in the pre- and post-conditions, but \emph{operation interfaces} instead.
Operation interfaces,\footnote{\url{https://mlir.llvm.org/docs/Interfaces/}} are introduced in MLIR to group similarly behaving operations, such as operations that perform a memory allocation or have a specific side-effect.

\paragraph{Advanced Pre- and Post-Condition}
Oftentimes, pre- and post-conditions of a transformation require more detail than purely demanding the presence of certain operations.
The Transform dialect supports advanced pre- and post-conditions via integration with the declarative IR Definition Language (IRDL)~\cite{irdl}.
IRDL is primarily used to declaratively specify operations of dialects with their types and invariants.
\Cref{fig:memref_subview_def} shows the IRDL definition of the \texttt{memref.subview} operation, that converts one memory reference type to another type which represents a reduced-size view of the original memory.

To specify advanced transform pre- and post-conditions, we leverage IRDL's capability to further constrain operations and types for \emph{existing} operations, without changing their definition.

\begin{figure}
\centering

\begin{lstlisting}[style=IRDL, escapeinside={|}{|}]
Dialect memref {
Operation subview|\colorbox{yellow!30}{\strut .constr}| {
  Attributes(
    static_offsets: Variadic<!indexAttr>,
    static_sizes:   Variadic<!indexAttr>,
    static_strides: Variadic<!indexAttr>)
  Operands(
    input:    !memrefType,
    offset:   Variadic<!index|\colorbox{yellow!30}{ , 0}\vspace{-0.3em}|>,
    sizes:    Variadic<!index|\colorbox{yellow!30}{ , 0}\vspace{-0.3em}|>,
    strides:  Variadic<!index|\colorbox{yellow!30}{ , 0}\vspace{-0.3em}|>)
  Results(view: !memrefType)
  CPPConstraint "checkMemrefConstraints()" }}
\end{lstlisting}

\caption{IRDL definition of the \texttt{memref.subview} operation. Highlighted parts are added in a copy of definition that expresses the post-condition of the \texttt{transform.expand\_strided\_metadata} transformation.}
\label{fig:memref_subview_def}
\end{figure}

The \texttt{expand\_strided\_metadata} transformation, shown in \Cref{fig:memref_constraints}, modifies \texttt{memref} dialect operations so that complex strided address computations are factored out and the remaining accesses are akin to trivial flat pointers~\cite{linalg}.
These simplified accesses are characterized by trivial \texttt{subview}s where the access offsets, sizes and strides are empty.
\Cref{fig:memref_subview_def} with highlights shows the pseudo operation that we introduce to represent this constrained, where the \texttt{offset}, \text{sizes} and \texttt{stride} operands of a \texttt{subview} are guaranteed to have cardinality zero.
Note, that we only use this IRDL definition to be able to specify the post-condition in \Cref{fig:memref_constraints} and that we \emph{do not} actually introduce a new operation.

\begin{figure}[t]
\centering
\begin{lstlisting}[escapeinside={!}{!}, style=transform]
transform.expand_strided_metadata(input: {memref.!*!}) ->
  (result: {memref.subview.constr, 
    memref.extract_strided_metadata.constr,
    memref.extract_aligned_pointer_as_index.constr,
    memref.reinterpret_cast.constr,
    affine.min, affine.apply, arith.constant_index})
\end{lstlisting}
\caption{constraints on input and output handles of \texttt{transform.expand\_strided\_metadata}}
\label{fig:memref_constraints}
\end{figure}

\paragraph{Checking Pre- and Post-Conditions Dynamically}
Most existing MLIR transformations do not declare their pre- and post-conditions explicitly.
To help with the process of adding such declarations, we leverage IRDL's capability to automatically generate constraint verifiers. These verifiers can be used to dynamically check pre- and post-conditions.

These dynamic checks are performed while transforming a concrete input program and are also useful even when pre- and post-conditions have been specified explicitly -- and have been checked with our static tool.
We do this, as we can not check if the specified pre- and post-conditions are actually accurate specifications for the concrete transformation implementations usually written in C++.
Therefore, the dynamic checking can serve as an additional tool to detect bugs in transformations.

\subsection{The Transform IR}
\label{sec:transform_ir_benefits}
The implementation as a dialect in MLIR
enables multiple usage scenarios, where users either write Transform scripts directly in the MLIR format, or alternatively use one of the multiple alternative MLIR frontends, including Julia~\cite{juliaMLIRFrontend} or Python~\cite{levental2023nelli}.

Exposing the Transform script as IR also means that we can use compiler analyses and transformations on the Transform script itself.
This can be used in multiple scenarios.

\paragraph{Static analysis of handle invalidation}
Since transforms are just MLIR operations, static analysis for handle invalidation is readily available as of-the-shelf ``use after free'' dataflow analysis. It suffices to express handle definition as an ``allocate'' side effect and handle invalidation as a ``free'' effect on some notional memory location, and to express handles pointing to the same or nested payload operations as (partial) aliasing. \emph{This showcases the benefit of expressing Transform scripts as regular compiler IR.}

\paragraph{Composition, simplification and constant propagation}
Named macros described in \Cref{sec:extensibility} are function-like objects that can be easily processed by existing function transformations. Indeed they may be implemented by simply calling the inliner pass and instructing it to always inline functions. Since macros don't support recursion, which is itself verified by checking for cycles in their call graph, inlining is always possible.

This in turn enables other simplification using MLIR peephole optimizer and constant folder. Similarly to any operation, transform operations can define local simplification rules, e.g., unrolling by 1 or tiling by 0 are noops, so is tiling by a larger size than the previously applied tiling. Constant parameters and typing information can be propagated through the script enabling further simplification using the pre-existing simplification driver. Note that performing this on the Transform IR removes the need for the transform to be applied to the payload, most often saving on compile-time for advanced transformations requiring expensive analyses.

\paragraph{Automatically configuring transformation pipelines via introspection}
Some transformations are meaningful at different levels of abstraction.
Automatic differentiation (AD), a process crucial for machine learning training, is such a transformation and is often implemented as a compiler pass~\cite{enzyme}.
AD produces instructions computing a derivative of a given variable. The full derivative is systematically a sum of partial derivatives with respect to different values. However, the notion of a sum is ambiguous as there are multiple kinds of additions, implemented in various MLIR dialects. The AD pass needs to create ``add'' instructions of the right kind for IR to remain compatible with the rest of the transformation pipeline. This in turn requires the AD pass to know its place in the pass pipeline and adjust accordingly.
In the JAX framework~\cite{jax}, the IR is progressively rewritten from StableHLO\footnote{\url{https://openxla.org/stablehlo}} to MHLO\footnote{\url{https://www.tensorflow.org/mlir/hlo_ops}}, to Arithmetic, and LLVM dialects at various stages, each of which defines an ``add'' operation. Depending on when the AD pass is scheduled, it must produce the operation from the corresponding dialect.
While the pass could analyze the IR to understand which ``stage'' it is in, such an analysis could be imprecise since multiple dialects can co-exist in the same translation unit and maintaining fine-grained control is important.

\begin{figure}
    \centering
    \includegraphics[width=0.48\textwidth]{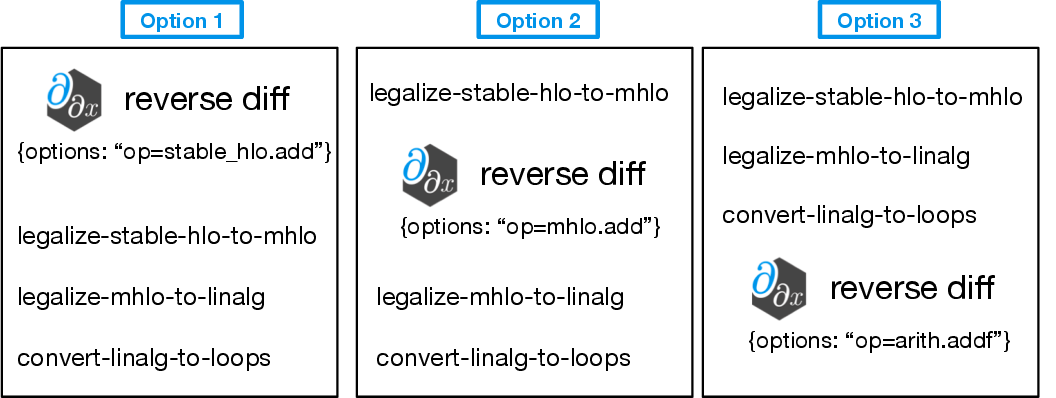}
    \caption{Three alternative options for when to perform the automatic differentiation transformation.
    We use a transform to introspect the pipeline and automatically infer the appropriate transformation option.
    }
    \label{fig:enzyme_options}
\end{figure}

Using a Transform script, we can precisely control the abstraction level at which we want to perform AD, as shown in \Cref{fig:enzyme_options}.
However, we must manually configure the automatic differentiation pass~\cite{enzyme} that operates on MLIR and has been parameterized by the kind of addition operation to emit.
To avoid manually specifying this unnecessary detail, we introspect the Transform script to infer the parameter based on the position in the script.
This is implemented as an ordinary compiler transformation performing a simple automatic traversal over the Transform IR.

\subsection{Summary}
In this section, we introduced the Transform dialect for composing and controlling compiler transformations.
We discussed various features, including handle invalidation, extensiblity, ensuring composability via pre- and post-conditions, and the representation of Transform scripts as MLIR IR.
Next, we are evaluating transform using a number of case studies.

\section{Evaluation}
\label{sec:case_studies}

We explore 5 case studies with different compiler scenarios.
We investigate the overhead of the Transform dialect and highlight the usefulness of pre- and post-conditions for building robust compiler pipelines, before showcasing the fine-grained control helpful for detecting performance problems, generating high-performance code, and automatically explore compiler optimizations.

\subsection{Case Study 1: Expressing arbitrary pass pipelines as Transform scripts}
\label{sec:arbitrary_pass_pipelines}
First, we want to establish that we can represent traditional pass pipelines with the Transform dialect and measure its overheads due to the interpretation at compile time.

Prior work on scheduling languages including Halide~\cite{halide} and ELEVATE~\cite{rise} validated the effectiveness of utilizing schedule-based compilation strategies for single kernel optimization.
However, these approaches are either domain-specific or have not been integrated into larger realistic code generation settings.
We explore the feasibility of scaling such schedule-based compilation approaches to modern compiler infrastructures, focusing on large-scale programs such as full machine learning models.
In this case study, we measure the overhead of using our Transform scripts that are interpreted at compile time and compare this to the traditional way in MLIR to control compiler transformations: pass pipelines, that are specified as command line arguments to the compiler.

\begin{table}
{\small
\begin{tabular}{lrrr}
\toprule
\textbf{Model} & \textbf{\# Ops} & \multicolumn{2}{c}{\textbf{Compile Time (ms)}} \\
 & & \textbf{MLIR} & \textbf{Transform} \\\midrule
Squeezenet~\cite{iandola2016squeezenet} & 126   & 16.6 & 16.9 \\
GPT-2~\cite{radford2019language} & 2861  & 185.4 & 190.0 \\
Mobile BERT~\cite{sun2020mobilebert} &4134  & 316.7 & 317.7 \\
Whisper (decoder only)~\cite{pmlr-v202-radford23a} & 847  & 457.5 & 462.3 \\
BERT-base-uncased~\cite{DBLP:journals/corr/abs-1810-04805} & 1182   & 1315.3 & 1348.6 \\
\bottomrule
\end{tabular}
}
\caption{\label{tab:pipeline_models} ML models converted to TOSA from TensorFlow using
\texttt{\scriptsize flatbuffer\_translate -tflite-flatbuffer-to-mlir} and \texttt{\scriptsize tf-opt --tfl-to-tosa-pipeline}. The use of the Transform dialect introduces $\leq 2.6\%$ compile time overhead.}
\end{table}

We evaluate the overhead of using the Transform dialect using five machine learning models listed in \Cref{tab:pipeline_models} and implemented with the MLIR-based TensorFlow compiler ecosystem.
The following pass pipeline from the MLIR Tensor Operation Set Architecture (TOSA) dialect\footnote{\url{https://www.mlplatform.org/tosa}} to the Linalg dialect~\cite{linalg} is described in \cite{10.1007/978-3-031-13188-2_19}:
\noindent
{\makeatletter
\patchcmd{\@verbatim}
  {\verbatim@font}
  {\verbatim@font\fontsize{7}{7.5}\selectfont}
  {}{}
\makeatother
\begin{verbatim}
mlir-opt --pass-pipeline=builtin.module(
  func.func(tosa-optional-decompositions),
  canonicalize,
  func.func(tosa-infer-shapes,tosa-make-broadcastable,
            tosa-to-linalg-named),
  canonicalize,
  func.func(tosa-layerwise-constant-fold,
            tosa-make-broadcastable),
  tosa-validate,
  func.func(tosa-to-linalg,tosa-to-arith,tosa-to-tensor),
  linalg-fuse-elementwise-ops,
  one-shot-bufferize)
\end{verbatim}}
\noindent

To obtain a Transform script describing the identical compilation flow, we modified MLIR to automatically create a Transform script of a pass pipeline that uses the generic \texttt{transform.apply\_registered\_pass} transform operation to invoke MLIR passes.
We compare MLIR's built-in pass manager system with the Transform dialect by running the identical pass pipelines, which is the worst-case scenario for the Transform dialect, as we do not make use of any of its useful features to precisely control the compilation process -- instead we are measuring the pure overhead.

Our measurements in \Cref{tab:pipeline_models} and \Cref{fig:compile_time_comparison} show, that the Transform dialect introduces very little compile time overhead, up to $2.6\%$ compared to the default pass pipeline.
Most of the passes used in these compilation flows are relatively simple, we expect even less compile time overhead when more complex passes such as register allocation are invoked.

\begin{figure}[t]
\centering
\scalebox{0.75}{%
\begin{tikzpicture}
\begin{loglogaxis}[
    xlabel={Compile Time MLIR (ms)},
    ylabel={Compile Time Transform (ms)},
    xmin=1, xmax=10000,
    ymin=1, ymax=10000,
    axis lines=center,
    axis equal image=true,
    grid=major,
]
\addplot[only marks, color=blue] coordinates {
    (0016.56433, 0016.91223) 
    (0457.5162, 0462.31785) 
    (1315.378, 1348.599) 
    (0185.4009, 0190.0676) 
    (0316.7464, 0317.70085) 
};

\addplot[domain=1:1000000, samples=2, dashed, color=red] {x};
\end{loglogaxis}
\end{tikzpicture}
}
\caption{Compile time comparison of different models using MLIR vs using }
\label{fig:compile_time_comparison}
\end{figure}

\subsection{Case Study 2: Building robust pipelines for lowering a soup of dialects}
In this case study, we are interested in highlighting the importance of pre- and post-conditions for building robust and flexible compilation pipelines in MLIR.

As we discussed in \Cref{sec:transform_dialect}, compiler transformations often have specific assumptions on their input, which are not explicitly expressed or checked.
Because of that, when exploring possible pass pipelines developers often encounter compiler errors, which are hard to relate to the underlying problem of an invalid pass order.

In this case study, we specifically investigate programs represented using the MLIR compiler infrastructure that are composed out of a mix of different dialects. For instance, arithmetics are represented using the \texttt{arith} dialect, indexing using the \texttt{index} dialect, memory using the \texttt{memref} dialect,

loops using the \texttt{scf} dialect, and functions using the \texttt{func} dialect. An example of a simple program that leverages all of these dialects is the following:
\begin{lstlisting}[style=MLIR,escapeinside={|}{|}]
func.func @chunkTo42(A: memref<64x64xf64>) {
  %chunk = memref.subview %A[|{\color{gray}/*offsets=*/}| 0, 0]    |{\color{gray}[/*sizes=*/}| 4, 4][|{\color{gray}/*strides=*/}| 1, 1] :
    memref<64x64xf64> to memref<4x4xf64, ...>
  %value = arith.constant 42.0
  scf.forall (%i, %j) = (0, 0) to (4, 4) {
    memref.store %value, %chunk[%i, %j]
  }
}
\end{lstlisting}

This function accepts a reference to a two-dimensional array in memory (\texttt{memref}) and creates a 4x4 rectangular \emph{view} of a part of it at the given offset and potentially with strides. All values in the view are then set to the value 42. Such (sub)views support access with local indexing without any additional index computations and modifications of the loop.

Lowering this simple IR for execution already requires running specific passes to progressively lower these dialects to LLVM. An example of a minimal pass pipeline is:
\begin{table*}
\centering
\begin{tabular}{l l l}
\hline
\textbf{Transform Operation}& \textbf{Pre-conditions} & \textbf{Post-conditions} \\
\hline
\Circled{1} convert-scf-to-cf           & $\{sc\hspace{-0.1em}f\hspace{-0.1em}.*\}         $ & $\{c\hspace{-0.1em}f\hspace{-0.1em}.\{branch, cond\_branch\}, arith.\{addi, cmpi, ...\},cast\}$ \\
\Circled{2} convert-arith-to-llvm       & $\{arith.*\}       $ & $\{llvm.\{add, f\hspace{-0.1em}add, bitcast, f\hspace{-0.1em}div, sdiv, udiv, ...\}, cast\}$ \\
\Circled{3} convert-cf-to-llvm          & $\{cf\hspace{-0.1em}.*\}          $ & $\{llvm\{f\hspace{-0.1em}unc, br, call, cond\_br, switch, unreachable\}, cast\}$ \\

\Circled{4} convert-func-to-llvm        & $\{f\hspace{-0.1em}unc.*\}        $ & $\{llvm.\{alloca, call, constant, f\hspace{-0.1em}unc, load, store, unde\hspace{-0.1em}f\hspace{-0.1em}, ...\},cast\}$ \\
\Circled{5} expand-strided-metadata     & $\{memre\hspace{-0.1em}f\hspace{-0.1em}.*\}$ & $\{memre\hspace{-0.1em}f\hspace{-0.1em}.\{subview.constr\},llvm.\{load,...\}, {\color{red}\textit{affine.apply}}\}$ \\
\Circled{6} finalize-memref-to-llvm     & $\{memre\hspace{-0.1em}f\hspace{-0.1em}.subview.constr\} $ & $\{llvm.\{add, alloca, br, call, constant, load, ptrtoint, ...\}, cast\}$ \\
\Circled{7} reconcile-unrealized-casts    & $\{cast\}$& $\{\}$ \\
\hline
\end{tabular}
\caption{Pre-/post-conditions conditions indicate payload operations removed/introduced by a transform, or additional constraints from \Cref{fig:memref_subview_def}. The \emph{affine.apply} operation potentially introduced by \protect\Circled{5} is not removed by any following pass. Thus the final IR is $\{llvm.*,{\textit{affine.apply}}\}$ and not only LLVM dialect.}
\label{fig:prePostConditions}
\end{table*}
\begin{itemize}
    \item[\Circled{1}] \texttt{convert-scf-to-cf}\\
    Lower structured control flow (\texttt{scf.forall}) to basic blocks and branching instructions.
    \item[\Circled{2}] \texttt{convert-arith-to-llvm}\\
    Lower arithmetic operations to their LLVM dialect counterparts.
    \item[\Circled{3}] \texttt{convert-cf-to-llvm}\\
    Lower control flow (e.g. \texttt{cf.br}) to LLVM counterparts.
    \item[\Circled{4}] \texttt{convert-func-to-llvm}\\
    Lower functions to an LLVM compatible format (e.g. return multiple values as a structure).
    \item[\Circled{5}] \texttt{expand-strided-metadata}\\
    Externalize non-trivial addressing from memrefs.
    \item[\Circled{6}] \texttt{finalize-memref-to-llvm}\\
    Lower trivially indexed memrefs to LLVM pointers.
    \item[\Circled{7}] \texttt{reconcile-unrealized-casts}\\
    Eliminate temporary type cast operations introduced by previous passes when possible.
\end{itemize}

Unfortunately, this pipeline fails as soon as we slightly change the input program by
having the view created at the non-zero offset provided as an additional function argument, \texttt{\%offset\,:\,index}.

This results in the following error: \emph{``failed to legalize operation 'builtin.unrealized\_conversion\_cast' that was explicitly marked illegal''}. 
This message does not point towards a solution, making the user resort to a painstaking inspection of the IR after each pass.

This error is due to an \texttt{affine.apply} operation that has been introduced during the \texttt{-expand-strided-metadata} pass. It is used to model the indexing behavior of the now slightly more complex \texttt{memref.subview} operation. The following pass \Circled{6} cannot lower the \texttt{affine.apply} operation accepting \texttt{index} types to LLVM dialect types, so it inserts type casts around it assuming it will be handled by a successor pass that would insert the reverse casts. In its absence, the final pass \Circled{7} cannot remove these type casts that do not cancel out and reports the error.

Even this close IR inspection may be surprising to the user since they did use the \texttt{affine} dialect, mostly used to represent constructs amenable to the polyhedral model~\cite{polyhedral,affine}, in the input program. Moreover, this dialect may be seen as operating on a conceptually higher level than memory address indexing due to overly simplistic interpretation of MLIR memory references as mere pointers, and thus unexpected to be produced by a \emph{lowering} pass.

This example illustrates the challenge to compose robust pass pipelines.

An ad-hoc solution to this problem is to simply add the \texttt{lower-affine} pass and other transitively required lowering passes after the \texttt{-expand-strided-metadata} pass to lower all created operations from the \texttt{affine} dialect, including another application of \Circled{2}.

In the Transform dialect, we address this issue using pre- and post-conditions encoded in the corresponding transform operations and types, as shown in \Cref{fig:prePostConditions}.
Pre-conditions indicate the payload operations that are expected in the input and will be removed, other operations won't be modified. Post-conditions indicate new payload operations that will be produced.

Given the final condition of only using the LLVM dialect, $\{llvm.*\}$, our static checking tool reports an error in this pipeline as it identifies that an \emph{affine.apply} operation produced by \Circled{5} will remain after the pipeline.

The pre- and post-conditions of the Transform dialect support users to develop robust lowering pipelines that are known to work for all possible inputs.

\subsection{Case Study 3: Debugging Performance Problematic Optimization Patterns}
Besides ensuring that pipelines are correctly producing code, performance engineers often also have to deal with counter-productive effects of program optimizations, chasing and eliminating them.

For instance, while introducing additional peephole optimization patterns for StableHLO that is a part of the Enzyme AD workflow, we observed that a combination of over 100 work-reducing and enabling transformations yielded counter-productive results in one of the LLMs we were trying to optimize, with up to 9\% overall performance penalty compared to the JAX/XLA baseline. However, these optimization patterns are designed either for obvious work reduction (e.g., not adding tensor elements produced by padding with zero) or to enable other transformations (permute tensor transpose to enable it to be folded into a matrix multiplication that supports transposed operands).

In order to identify which of the individual patterns was counter-productive, we attempted to perform binary search over the pattern set. Since the set of optimization patterns is expressed in C++, it required manual code modifications and up to 10 minutes per individual pattern for a fresh compilation, linking and packaging on a 4x24-core Xeon Platinum 8160 (Skylake SP) platform with 196 GB RAM using LLVM.\footnote{LLVM toolchain \texttt{3b5e7c83a6e} from Mar 14, 2024} While re-compilation time of a single file is negligible, linking and packaging a 5.4 GiB self-contained tool that includes parts of TensorFlow, LLVM and a Python interpreter (common for production-oriented “hermetic” builds) consumes most of the time: 31s for linking and 164s for compressed packaging. To alleviate this, we leveraged the Transform dialect support for pattern application by associating each pattern with a transform operation as follows.\footnote{
Exact commit hash omitted for double-blindness.

}

\begin{lstlisting}[style=transform]
transform.apply_patterns to %func {
  transform.pattern.add_of_zero_pad
  transform.pattern.negate_of_transpose
  transform.pattern.matmul_of_transpose
  // more patterns
}
\end{lstlisting}

This allowed us to only deploy the compiler once and to automate the binary search over the pattern set by simply removing parts of the pattern list in the Transform script, with each iteration of binary search taking up to 4 seconds for compilation of the model. Thanks to this, we identified the counter-productive transformation as ``fold reshape/transpose into full reduce’’. While the full additive reduction of a tensor into a scalar can be implemented effectively regardless of the tensor shape (assuming associativity of floating point addition, e.g., \texttt{-ffast-math}, as is common for ML workloads), removing the leading reshape/transpose operations strictly reduces work. However, in this case, this adversely affected the fusion heuristic in the back-end XLA compiler\footnote{https://openxla.org/xla} that produced larger, less cache-efficient fusion clusters.

This case study demonstrates the usefulness of flexible and fine-grained compiler control that enables quickly exploring the space of transformations, in this instance to detect a performance problem that affects only a single, but important, input program.

\subsection{Case Study 4: Fine-Grained Control of Performance Optimizations}
\label{sec:casestudy4}
Ultimately, we are interested in building compilers that generate high-performance code.
The Transform dialect enables fine-grained control to specify compiler transformations, e.g., here for a single loop nest of a layer of the ResNet-50 model~\cite{he2015deep}. 

In this case study, we discuss an existing alternative in the form of OpenMP pragmas for controlling compiler optimizations.
We show the limitations of OpenMP and where the Transform dialect allows to go further by controlling and composing arbitrary compiler transformations.

OpenMP provides directives to perform loop transformations, that can have significant performance benefits, such as parallelization or vectorization.
These transformations also include tiling, as shown in \Cref{lst:openmp_program}.
Such transformations are naturally equally available in the Transform dialect.
Unlike OpenMP, however, the Transform script is not written directly together with the computational program.
Instead, we explicitly match the ``for'' loop, as in the line~2 of \Cref{lst:batchmatmul_transform_script}, which we then tile in line~4.

Since the trip count of the loop along \texttt{i} (196) is not divisible by the corresponding tile size (32), tiling will result in conditionals being introduced into the loop body, or alternatively the last iterations are being peeled off.
Having explicit handles in the Transform dialect allow us to precisely control this behavior.
We first split the loop into the divisible part and the remainder in line~3 and gain a handle for both of these loops which are not visible in the original code.
We then unroll only the remainder loop, named \texttt{\%rest}, in line~9.
Using OpenMP, we only have a limited way of composing transformations, so applying a transformation to a nested loop resulting from tiling is not possible.

We measured the performance of the generated code optimized via OpenMP and Transform and observed almost identical performance with the OpenMP version having a median runtime of 0.48 seconds and the Transform version of 0.49 seconds.

\begin{figure}[t]
\centering
\begin{lstlisting}[style=OpenMP]
for (int b=0; b<6; b++) {
  #pragma omp tile sizes(32, 32)
  for(int i=0; i<196; i++) {
    for(int j=0; j<256; j++) {
      for(int k=0; k<2305; k++) {
        C[b,i,j] += A[b,i,k] * B[b,k,j];}}}}
\end{lstlisting}
\caption{OpenMP pragmas for tiling a single loop nest.}
\label{lst:openmp_program}
\end{figure}

While OpenMP is limited to perform a fixed set of loop optimizations with the Transform dialect we can go further.

We introduce a new transform replacing a small fixed-size matrix multiplication, such as that formed by inner loops after tiling, with a call to a microkernel library~\cite{libxsmm}.
This new specialized transform is easily added into the compiler using the MLIR plugin mechanism and mixed with other transformations, as shown in line~7 of \Cref{lst:batchmatmul_transform_script}.

We furthermore wrap this transformation into the \texttt{alternatives} construct in lines 6--8 as it may fail when the microkernel library doesn't have an implementation with required sizes.
Here we give no additional alternative, so if the replacement with a library call fails, the input code remains unchanged.
With the optimized microkernel, we achieve a significantly better performance of 0.017s, over 20 times faster than the tiled versions.
This small experiment showcases the additional possibilities the Transform dialect opens to integrate highly specialized optimizations, such as replacements with dedicated library calls, over more rigid alternatives such as OpenMP.

\begin{figure}[t]
\centering
\begin{lstlisting}[style=transform]
transform.named_sequence @transform_main(%module) {
  %i_loop = match.op {second} "scf.for" in %module
  %main, %rest = loop.split %i_loop div_by 32
  %tiled:2 = loop.tile %main
              {tile_sizes=[32, 32]}
  transform.alternatives {
    transform.to_library %tiled#2 "libxsmm"
  }, { }
  loop.unroll %rest {full}
}
\end{lstlisting}
\caption{Transform script performing in lines 2--5 the same optimization as the OpenMP version above.
In line 7, the script attempts to replace the nested loop code with a library call resulting in a significant performance win.}
\label{lst:batchmatmul_transform_script}
\end{figure}

\subsection{Case Study 5: Performance Exploration with State-of-the-Art Autotuning Methods}

\begin{figure}[t]
\centering
\begin{lstlisting}[style=transform]
transform.named_sequence @transform_main(%module) {
  %batch_loop = match.op {first} "scf.for" in %module
  %tiled:4 = loop.tile %batch_loop
              {tile_sizes=[tile0,tile1,tile2,tile3]} 
  transform.alternatives {
    transform.to_library %tiled#4 "libxsmm"
  }, { 
    transform.assert vect
    transform.vectorize %tiled#4
  }, { }
  loop.unroll %rest {full}
}
\end{lstlisting}
\caption{Transform script with the parametric tile sizes that are automatically chosen using an autotuning tool.}
\label{lst:parametric_batchmatmul_transform_script}
\end{figure}

In this case study, we demonstrate how we can use Transform scripts to quickly explore an optimization space.
For this, we use the state-of-the-art bayesian autotuning tool BACO~\cite{baco} to automatically explore the search space for tiling the loop nest shown in \Cref{sec:casestudy4} according to the tuning parameters and constraints shown in \Cref{lst:tuning_settings}.

\begin{figure}[t]
    \centering
    \includegraphics[width=0.35\textwidth]{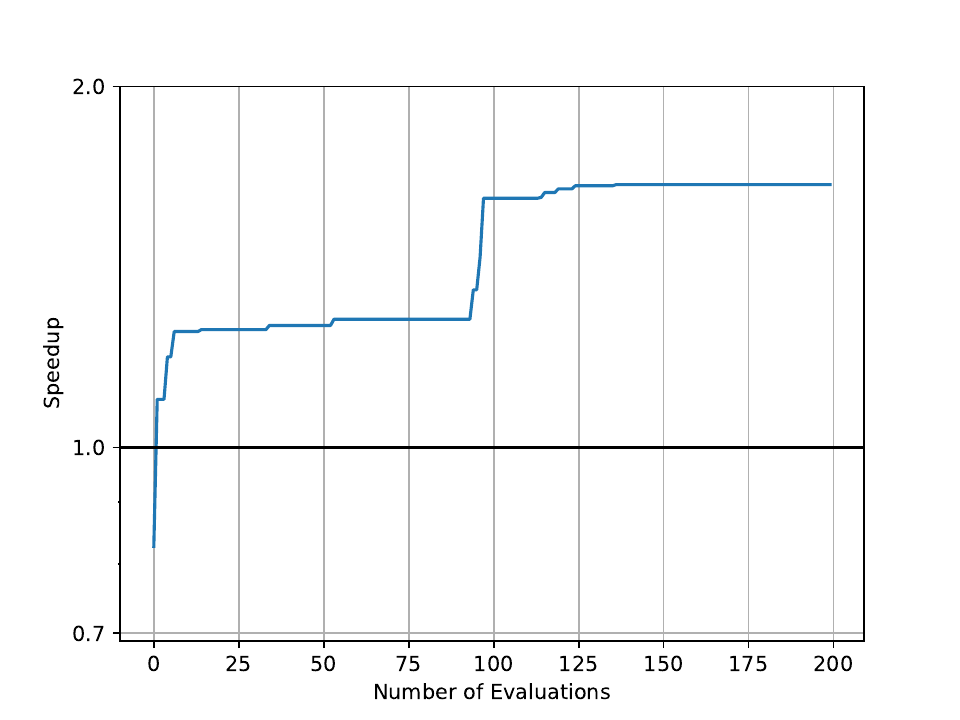}
    \caption{Performance evolution of batch matrix multiplication when searching for the best tile size parameters in the Transform script specifying the optimization.}
    \label{fig:my_label}
\end{figure}

\Cref{fig:my_label} shows the performance evolution of the optimization process.
The search gradually finds better and better values for the tile size parameters reaching a final speedup of 1.68.

\begin{figure}[t]
\centering
\begin{lstlisting}[style=transform]
tuning_parameters: [
  tile0: {range:[0, B], constraints:[B % tile0 == 0]}
  tile1: {range:[0, M], constraints:[M % tile1 == 0]}
  tile2: {range:[0, N], constraints:[N % tile2 == 0]}
  tile3: {range:[0, K], constraints:[K % tile3 == 0]}
  vect:  {range:[0, 1], constraints:[             where(tile3 % vector_size != 0, vect == 0)]} ]
\end{lstlisting}

\caption{Definition of the tuning parameters used in \Cref{lst:parametric_batchmatmul_transform_script} with constraints: tile sizes must divide their dimension and vectorization is disabled if the trip count of the inner-most loop is not divisible by the machine vector size.}
\label{lst:tuning_settings}
\end{figure}

This case study shows how easy it is to integrate Transform with autotuning and similar machine learning-driven exploration tools to quickly explore optimization spaces.

\section{Related Work}

\noindent\textbf{Controlling Compiler Optimizations.}
There are many alternative approaches to control compiler optimizations.
From manual annotations, such as OpenMP, over simple compiler flags, such as \texttt{-O3}, to more complex compiler flags, such as Polly's influencing heuristics \cite{grosser2012polly}.
These knobs can be used to search for the best performing combination of optimizations, e.g., tuning flags via Monte-Carlo search~\cite{DBLP:conf/pmbs-ws/KooBKWHH21} or machine learning methods~\cite{DBLP:conf/cgo/CumminsWGCAGJLT22}.
Exposing compiler heuristics to make them accessible is an active research area~\cite{DBLP:conf/cgo/SeekerCCFHL24}.

URUK~\cite{uruk} and Clay~\cite{clay} define directives for source-to-source polyhedral transformation.
Source matching and rewriting has also been proposed for MLIR~\cite{espindola2023source}.

The Transform dialect offers a significantly more fine-grained control of compiler transformations.
As our final case study showed this can also be combined with autotuning methods to explore the optimization space.

\smallskip
\noindent\textbf{Scheduling Languages.}
Halide~\cite{halide}, TVM~\cite{tvm}, TACO~\cite{taco}, and Rise~\cite{rise} are examples of domain-specific compilers that have scheduling languages allowing performance experts to control compiler optimizations.
Besides being domain-specific, the transformations exposed by the compilers are also fixed.
In contrast, the Transform dialect is an extensible scheduling language available to generic compilers.

ELEVATE~\cite{rise} is a scheduling language expressing compiler optimizations as compositions of formal rewrite rules.
It provides formal reasoning capabilities about compiler transformations~\cite{DBLP:journals/pacmpl/QinOGHKS24}.
While generic and extensible, ELEVATE needs optimizations expressed as formal rewrites, requiring computational programs to be pure to apply safely.
The Transform dialect enables composing and controlling of compiler transformations over a wider range of scenarios, in a pre-existing compiler framework.

Exo \cite{10.1145/3519939.3523446} makes scheduling languages a programmable extension to the compiler, by using algebraic rewrites within a core language instead of a specialized code generator template, and by generalizing Halide schedules to imperative code and relying on an SMT solver to verify side-effecting computations. Unlike the Transform dialect, Exo does not handle extensible intermediate representations (the core language is fixed), and thus cannot operate at the different levels of abstraction needed to operate on the pass pipeline itself, to reuse existing transformation passes, and does not decouple effecting transformations from their validation.

\section{Conclusion}
We presented and evaluated the MLIR Transform dialect.
It allows for fine-grained control of compiler optimizations while maintaining the extensible nature of MLIR.
Considering 5 case studies we have demonstrated its low overhead, and usefulness to define robust, flexible and high-performance compiler pipelines.

Its implementation in the most popular domain-specific compiler framework opens up many interesting applications to further exploit the already existing power of compilers in as many scenarios as possible.

\bibliographystyle{ACM-Reference-Format}
\bibliography{bibliography}


\begin{thebibliography}{38}


\ifx \showCODEN    \undefined \def \showCODEN     #1{\unskip}     \fi
\ifx \showDOI      \undefined \def \showDOI       #1{#1}\fi
\ifx \showISBNx    \undefined \def \showISBNx     #1{\unskip}     \fi
\ifx \showISBNxiii \undefined \def \showISBNxiii  #1{\unskip}     \fi
\ifx \showISSN     \undefined \def \showISSN      #1{\unskip}     \fi
\ifx \showLCCN     \undefined \def \showLCCN      #1{\unskip}     \fi
\ifx \shownote     \undefined \def \shownote      #1{#1}          \fi
\ifx \showarticletitle \undefined \def \showarticletitle #1{#1}   \fi
\ifx \showURL      \undefined \def \showURL       {\relax}        \fi
\providecommand\bibfield[2]{#2}
\providecommand\bibinfo[2]{#2}
\providecommand\natexlab[1]{#1}
\providecommand\showeprint[2][]{arXiv:#2}

\bibitem[Appel(1998)]%
        {appel1998ssa}
\bibfield{author}{\bibinfo{person}{Andrew~W Appel}.}
  \bibinfo{year}{1998}\natexlab{}.
\newblock \showarticletitle{{SSA} is functional programming}.
\newblock \bibinfo{journal}{\emph{Acm Sigplan Notices}} \bibinfo{volume}{33},
  \bibinfo{number}{4} (\bibinfo{year}{1998}), \bibinfo{pages}{17--20}.
\newblock


\bibitem[Bagn{\`e}res et~al\mbox{.}(2016)]%
        {clay}
\bibfield{author}{\bibinfo{person}{L{\'e}na{\"\i}c Bagn{\`e}res},
  \bibinfo{person}{Oleksandr Zinenko}, \bibinfo{person}{St{\'e}phane Huot},
  {and} \bibinfo{person}{C{\'e}dric Bastoul}.} \bibinfo{year}{2016}\natexlab{}.
\newblock \showarticletitle{Opening polyhedral compiler's black box}. In
  \bibinfo{booktitle}{\emph{Proceedings of the 2016 International Symposium on
  Code Generation and Optimization}}. \bibinfo{pages}{128--138}.
\newblock


\bibitem[Bang et~al\mbox{.}(2022)]%
        {10.1007/978-3-031-13188-2_19}
\bibfield{author}{\bibinfo{person}{Seongwon Bang}, \bibinfo{person}{Seunghyeon
  Nam}, \bibinfo{person}{Inwhan Chun}, \bibinfo{person}{Ho~Young Jhoo}, {and}
  \bibinfo{person}{Juneyoung Lee}.} \bibinfo{year}{2022}\natexlab{}.
\newblock \showarticletitle{SMT-Based Translation Validation for Machine
  Learning Compiler}. In \bibinfo{booktitle}{\emph{Computer Aided
  Verification}}, \bibfield{editor}{\bibinfo{person}{Sharon Shoham} {and}
  \bibinfo{person}{Yakir Vizel}} (Eds.). \bibinfo{publisher}{Springer
  International Publishing}, \bibinfo{address}{Cham},
  \bibinfo{pages}{386--407}.
\newblock
\showISBNx{978-3-031-13188-2}


\bibitem[Bondhugula(2020)]%
        {affine}
\bibfield{author}{\bibinfo{person}{Uday Bondhugula}.}
  \bibinfo{year}{2020}\natexlab{}.
\newblock \bibinfo{title}{High Performance Code Generation in MLIR: An Early
  Case Study with GEMM}.
\newblock
\newblock
\showeprint[arxiv]{2003.00532}~[cs.PF]


\bibitem[Chen et~al\mbox{.}(2018)]%
        {tvm}
\bibfield{author}{\bibinfo{person}{Tianqi Chen}, \bibinfo{person}{Thierry
  Moreau}, \bibinfo{person}{Ziheng Jiang}, \bibinfo{person}{Lianmin Zheng},
  \bibinfo{person}{Eddie Yan}, \bibinfo{person}{Haichen Shen},
  \bibinfo{person}{Meghan Cowan}, \bibinfo{person}{Leyuan Wang},
  \bibinfo{person}{Yuwei Hu}, \bibinfo{person}{Luis Ceze}, {et~al\mbox{.}}}
  \bibinfo{year}{2018}\natexlab{}.
\newblock \showarticletitle{{TVM}: An automated {End-to-End} optimizing
  compiler for deep learning}. In \bibinfo{booktitle}{\emph{13th USENIX
  Symposium on Operating Systems Design and Implementation (OSDI 18)}}.
  \bibinfo{pages}{578--594}.
\newblock


\bibitem[Cohen et~al\mbox{.}(2005)]%
        {uruk}
\bibfield{author}{\bibinfo{person}{Albert Cohen}, \bibinfo{person}{Marc
  Sigler}, \bibinfo{person}{Sylvain Girbal}, \bibinfo{person}{Olivier Temam},
  \bibinfo{person}{David Parello}, {and} \bibinfo{person}{Nicolas Vasilache}.}
  \bibinfo{year}{2005}\natexlab{}.
\newblock \showarticletitle{Facilitating the search for compositions of program
  transformations}. In \bibinfo{booktitle}{\emph{Proceedings of the 19th annual
  international conference on Supercomputing}}. \bibinfo{pages}{151--160}.
\newblock


\bibitem[Cummins et~al\mbox{.}(2022)]%
        {DBLP:conf/cgo/CumminsWGCAGJLT22}
\bibfield{author}{\bibinfo{person}{Chris Cummins}, \bibinfo{person}{Bram
  Wasti}, \bibinfo{person}{Jiadong Guo}, \bibinfo{person}{Brandon Cui},
  \bibinfo{person}{Jason Ansel}, \bibinfo{person}{Sahir Gomez},
  \bibinfo{person}{Somya Jain}, \bibinfo{person}{Jia Liu},
  \bibinfo{person}{Olivier Teytaud}, \bibinfo{person}{Benoit Steiner},
  \bibinfo{person}{Yuandong Tian}, {and} \bibinfo{person}{Hugh Leather}.}
  \bibinfo{year}{2022}\natexlab{}.
\newblock \showarticletitle{CompilerGym: Robust, Performant Compiler
  Optimization Environments for {AI} Research}. In
  \bibinfo{booktitle}{\emph{{IEEE/ACM} International Symposium on Code
  Generation and Optimization, {CGO} 2022, Seoul, Korea, Republic of, April
  2-6, 2022}}, \bibfield{editor}{\bibinfo{person}{Jae~W. Lee},
  \bibinfo{person}{Sebastian Hack}, {and} \bibinfo{person}{Tatiana Shpeisman}}
  (Eds.). \bibinfo{publisher}{{IEEE}}, \bibinfo{pages}{92--105}.
\newblock
\urldef\tempurl%
\url{https://doi.org/10.1109/CGO53902.2022.9741258}
\showDOI{\tempurl}


\bibitem[Devlin et~al\mbox{.}(2018)]%
        {DBLP:journals/corr/abs-1810-04805}
\bibfield{author}{\bibinfo{person}{Jacob Devlin}, \bibinfo{person}{Ming{-}Wei
  Chang}, \bibinfo{person}{Kenton Lee}, {and} \bibinfo{person}{Kristina
  Toutanova}.} \bibinfo{year}{2018}\natexlab{}.
\newblock \showarticletitle{{BERT:} Pre-training of Deep Bidirectional
  Transformers for Language Understanding}.
\newblock \bibinfo{journal}{\emph{CoRR}}  \bibinfo{volume}{abs/1810.04805}
  (\bibinfo{year}{2018}).
\newblock
\showeprint[arxiv]{1810.04805}
\urldef\tempurl%
\url{http://arxiv.org/abs/1810.04805}
\showURL{%
\tempurl}


\bibitem[Espindola et~al\mbox{.}(2023)]%
        {espindola2023source}
\bibfield{author}{\bibinfo{person}{Vinicius Espindola},
  \bibinfo{person}{Luciano Zago}, \bibinfo{person}{Herv{\'e} Yviquel}, {and}
  \bibinfo{person}{Guido Araujo}.} \bibinfo{year}{2023}\natexlab{}.
\newblock \showarticletitle{Source matching and rewriting for MLIR using
  string-based automata}.
\newblock \bibinfo{journal}{\emph{ACM Transactions on Architecture and Code
  Optimization}} \bibinfo{volume}{20}, \bibinfo{number}{2}
  (\bibinfo{year}{2023}), \bibinfo{pages}{1--26}.
\newblock


\bibitem[Feautrier and Lengauer(2011)]%
        {polyhedral}
\bibfield{author}{\bibinfo{person}{Paul Feautrier} {and}
  \bibinfo{person}{Christian Lengauer}.} \bibinfo{year}{2011}\natexlab{}.
\newblock \bibinfo{booktitle}{\emph{Polyhedron Model}}.
\newblock \bibinfo{publisher}{Springer US}, \bibinfo{address}{Boston, MA},
  \bibinfo{pages}{1581--1592}.
\newblock
\showISBNx{978-0-387-09766-4}
\urldef\tempurl%
\url{https://doi.org/10.1007/978-0-387-09766-4_502}
\showDOI{\tempurl}


\bibitem[Fehr et~al\mbox{.}(2022)]%
        {irdl}
\bibfield{author}{\bibinfo{person}{Mathieu Fehr}, \bibinfo{person}{Jeff Niu},
  \bibinfo{person}{River Riddle}, \bibinfo{person}{Mehdi Amini},
  \bibinfo{person}{Zhendong Su}, {and} \bibinfo{person}{Tobias Grosser}.}
  \bibinfo{year}{2022}\natexlab{}.
\newblock \showarticletitle{IRDL: an IR definition language for SSA compilers}.
  In \bibinfo{booktitle}{\emph{Proceedings of the 43rd ACM SIGPLAN
  International Conference on Programming Language Design and Implementation}}
  (San Diego, CA, USA) \emph{(\bibinfo{series}{PLDI 2022})}.
  \bibinfo{publisher}{Association for Computing Machinery},
  \bibinfo{address}{New York, NY, USA}, \bibinfo{pages}{199–212}.
\newblock
\showISBNx{9781450392655}
\urldef\tempurl%
\url{https://doi.org/10.1145/3519939.3523700}
\showDOI{\tempurl}


\bibitem[Frostig et~al\mbox{.}(2018)]%
        {jax}
\bibfield{author}{\bibinfo{person}{Roy Frostig}, \bibinfo{person}{Matthew~James
  Johnson}, {and} \bibinfo{person}{Chris Leary}.}
  \bibinfo{year}{2018}\natexlab{}.
\newblock \showarticletitle{Compiling machine learning programs via high-level
  tracing}.
\newblock \bibinfo{journal}{\emph{Systems for Machine Learning}}
  \bibinfo{volume}{4}, \bibinfo{number}{9} (\bibinfo{year}{2018}).
\newblock


\bibitem[Grosser et~al\mbox{.}(2012)]%
        {grosser2012polly}
\bibfield{author}{\bibinfo{person}{Tobias Grosser}, \bibinfo{person}{Armin
  Groesslinger}, {and} \bibinfo{person}{Christian Lengauer}.}
  \bibinfo{year}{2012}\natexlab{}.
\newblock \showarticletitle{Polly—performing polyhedral optimizations on a
  low-level intermediate representation}.
\newblock \bibinfo{journal}{\emph{Parallel Processing Letters}}
  \bibinfo{volume}{22}, \bibinfo{number}{04} (\bibinfo{year}{2012}),
  \bibinfo{pages}{1250010}.
\newblock


\bibitem[Hagedorn et~al\mbox{.}(2020)]%
        {rise}
\bibfield{author}{\bibinfo{person}{Bastian Hagedorn}, \bibinfo{person}{Johannes
  Lenfers}, \bibinfo{person}{Thomas Koehler}, \bibinfo{person}{Xueying Qin},
  \bibinfo{person}{Sergei Gorlatch}, {and} \bibinfo{person}{Michel Steuwer}.}
  \bibinfo{year}{2020}\natexlab{}.
\newblock \showarticletitle{Achieving high-performance the functional way: a
  functional pearl on expressing high-performance optimizations as rewrite
  strategies}.
\newblock \bibinfo{journal}{\emph{Proc. {ACM} Program. Lang.}}
  \bibinfo{volume}{4}, \bibinfo{number}{{ICFP}} (\bibinfo{year}{2020}),
  \bibinfo{pages}{92:1--92:29}.
\newblock
\urldef\tempurl%
\url{https://doi.org/10.1145/3408974}
\showDOI{\tempurl}


\bibitem[He et~al\mbox{.}(2015)]%
        {he2015deep}
\bibfield{author}{\bibinfo{person}{Kaiming He}, \bibinfo{person}{Xiangyu
  Zhang}, \bibinfo{person}{Shaoqing Ren}, {and} \bibinfo{person}{Jian Sun}.}
  \bibinfo{year}{2015}\natexlab{}.
\newblock \bibinfo{title}{Deep Residual Learning for Image Recognition}.
\newblock
\newblock
\showeprint[arxiv]{1512.03385}~[cs.CV]


\bibitem[Heinecke et~al\mbox{.}(2016)]%
        {libxsmm}
\bibfield{author}{\bibinfo{person}{Alexander Heinecke}, \bibinfo{person}{Greg
  Henry}, \bibinfo{person}{Maxwell Hutchinson}, {and} \bibinfo{person}{Hans
  Pabst}.} \bibinfo{year}{2016}\natexlab{}.
\newblock \showarticletitle{LIBXSMM: accelerating small matrix multiplications
  by runtime code generation}. In \bibinfo{booktitle}{\emph{SC'16: Proceedings
  of the International Conference for High Performance Computing, Networking,
  Storage and Analysis}}. IEEE, \bibinfo{pages}{981--991}.
\newblock


\bibitem[Hellsten et~al\mbox{.}(2024)]%
        {baco}
\bibfield{author}{\bibinfo{person}{Erik~Orm Hellsten}, \bibinfo{person}{Artur
  Souza}, \bibinfo{person}{Johannes Lenfers}, \bibinfo{person}{Rubens
  Lacouture}, \bibinfo{person}{Olivia Hsu}, \bibinfo{person}{Adel Ejjeh},
  \bibinfo{person}{Fredrik Kjolstad}, \bibinfo{person}{Michel Steuwer},
  \bibinfo{person}{Kunle Olukotun}, {and} \bibinfo{person}{Luigi Nardi}.}
  \bibinfo{year}{2024}\natexlab{}.
\newblock \showarticletitle{BaCO: A Fast and Portable Bayesian Compiler
  Optimization Framework}. In \bibinfo{booktitle}{\emph{Proceedings of the 28th
  ACM International Conference on Architectural Support for Programming
  Languages and Operating Systems, Volume 4}} (Vancouver, BC, Canada)
  \emph{(\bibinfo{series}{ASPLOS '23})}. \bibinfo{publisher}{Association for
  Computing Machinery}, \bibinfo{address}{New York, NY, USA},
  \bibinfo{pages}{19–42}.
\newblock
\showISBNx{9798400703942}
\urldef\tempurl%
\url{https://doi.org/10.1145/3623278.3624770}
\showDOI{\tempurl}


\bibitem[Iandola et~al\mbox{.}(2016)]%
        {iandola2016squeezenet}
\bibfield{author}{\bibinfo{person}{Forrest~N. Iandola}, \bibinfo{person}{Song
  Han}, \bibinfo{person}{Matthew~W. Moskewicz}, \bibinfo{person}{Khalid
  Ashraf}, \bibinfo{person}{William~J. Dally}, {and} \bibinfo{person}{Kurt
  Keutzer}.} \bibinfo{year}{2016}\natexlab{}.
\newblock \bibinfo{title}{SqueezeNet: AlexNet-level accuracy with 50x fewer
  parameters and <0.5MB model size}.
\newblock
\newblock
\showeprint[arxiv]{1602.07360}~[cs.CV]


\bibitem[Ikarashi et~al\mbox{.}(2022)]%
        {10.1145/3519939.3523446}
\bibfield{author}{\bibinfo{person}{Yuka Ikarashi},
  \bibinfo{person}{Gilbert~Louis Bernstein}, \bibinfo{person}{Alex Reinking},
  \bibinfo{person}{Hasan Genc}, {and} \bibinfo{person}{Jonathan Ragan-Kelley}.}
  \bibinfo{year}{2022}\natexlab{}.
\newblock \showarticletitle{Exocompilation for productive programming of
  hardware accelerators}. In \bibinfo{booktitle}{\emph{Proceedings of the 43rd
  ACM SIGPLAN International Conference on Programming Language Design and
  Implementation}} (San Diego, CA, USA) \emph{(\bibinfo{series}{PLDI 2022})}.
  \bibinfo{publisher}{Association for Computing Machinery},
  \bibinfo{address}{New York, NY, USA}, \bibinfo{pages}{703–718}.
\newblock
\showISBNx{9781450392655}
\urldef\tempurl%
\url{https://doi.org/10.1145/3519939.3523446}
\showDOI{\tempurl}


\bibitem[Kjolstad et~al\mbox{.}(2017)]%
        {taco}
\bibfield{author}{\bibinfo{person}{Fredrik Kjolstad}, \bibinfo{person}{Shoaib
  Kamil}, \bibinfo{person}{Stephen Chou}, \bibinfo{person}{David Lugato}, {and}
  \bibinfo{person}{Saman~P. Amarasinghe}.} \bibinfo{year}{2017}\natexlab{}.
\newblock \showarticletitle{The tensor algebra compiler}.
\newblock \bibinfo{journal}{\emph{Proc. {ACM} Program. Lang.}}
  \bibinfo{volume}{1}, \bibinfo{number}{{OOPSLA}} (\bibinfo{year}{2017}),
  \bibinfo{pages}{77:1--77:29}.
\newblock
\urldef\tempurl%
\url{https://doi.org/10.1145/3133901}
\showDOI{\tempurl}


\bibitem[Koo et~al\mbox{.}(2021)]%
        {DBLP:conf/pmbs-ws/KooBKWHH21}
\bibfield{author}{\bibinfo{person}{Jaehoon Koo}, \bibinfo{person}{Prasanna
  Balaprakash}, \bibinfo{person}{Michael Kruse}, \bibinfo{person}{Xingfu Wu},
  \bibinfo{person}{Paul~D. Hovland}, {and} \bibinfo{person}{Mary~W. Hall}.}
  \bibinfo{year}{2021}\natexlab{}.
\newblock \showarticletitle{Customized Monte Carlo Tree Search for LLVM/Polly's
  Composable Loop Optimization Transformations}. In
  \bibinfo{booktitle}{\emph{2021 International Workshop on Performance
  Modeling, Benchmarking and Simulation of High Performance Computer Systems
  {(PMBS} 2021), St. Louis, MO, USA, November 15, 2021}}.
  \bibinfo{publisher}{{IEEE}}, \bibinfo{pages}{82--93}.
\newblock
\urldef\tempurl%
\url{https://doi.org/10.1109/PMBS54543.2021.00015}
\showDOI{\tempurl}


\bibitem[Lattner and Adve(2004)]%
        {llvm}
\bibfield{author}{\bibinfo{person}{Chris Lattner} {and}
  \bibinfo{person}{Vikram~S. Adve}.} \bibinfo{year}{2004}\natexlab{}.
\newblock \showarticletitle{{LLVM:} {A} Compilation Framework for Lifelong
  Program Analysis {\&} Transformation}. In \bibinfo{booktitle}{\emph{2nd
  {IEEE} / {ACM} International Symposium on Code Generation and Optimization
  {(CGO} 2004), 20-24 March 2004, San Jose, CA, {USA}}}.
  \bibinfo{publisher}{{IEEE} Computer Society}, \bibinfo{pages}{75--88}.
\newblock
\urldef\tempurl%
\url{https://doi.org/10.1109/CGO.2004.1281665}
\showDOI{\tempurl}


\bibitem[Lattner et~al\mbox{.}(2021)]%
        {lattner2020mlir}
\bibfield{author}{\bibinfo{person}{Chris Lattner}, \bibinfo{person}{Mehdi
  Amini}, \bibinfo{person}{Uday Bondhugula}, \bibinfo{person}{Albert Cohen},
  \bibinfo{person}{Andy Davis}, \bibinfo{person}{Jacques Pienaar},
  \bibinfo{person}{River Riddle}, \bibinfo{person}{Tatiana Shpeisman},
  \bibinfo{person}{Nicolas Vasilache}, {and} \bibinfo{person}{Oleksandr
  Zinenko}.} \bibinfo{year}{2021}\natexlab{}.
\newblock \showarticletitle{{MLIR}: Scaling Compiler Infrastructure for Domain
  Specific Computation}. In \bibinfo{booktitle}{\emph{{CGO}}}.
\newblock


\bibitem[Levental et~al\mbox{.}(2023)]%
        {levental2023nelli}
\bibfield{author}{\bibinfo{person}{Maksim Levental}, \bibinfo{person}{Alok
  Kamatar}, \bibinfo{person}{Ryan Chard}, \bibinfo{person}{Kyle Chard}, {and}
  \bibinfo{person}{Ian Foster}.} \bibinfo{year}{2023}\natexlab{}.
\newblock \bibinfo{title}{nelli: a lightweight frontend for MLIR}.
\newblock
\newblock
\showeprint[arxiv]{2307.16080}~[cs.PL]


\bibitem[McKeeman(1965)]%
        {peephole}
\bibfield{author}{\bibinfo{person}{William~M. McKeeman}.}
  \bibinfo{year}{1965}\natexlab{}.
\newblock \showarticletitle{Peephole optimization}.
\newblock \bibinfo{journal}{\emph{Commun. {ACM}}} \bibinfo{volume}{8},
  \bibinfo{number}{7} (\bibinfo{year}{1965}), \bibinfo{pages}{443--444}.
\newblock
\urldef\tempurl%
\url{https://doi.org/10.1145/364995.365000}
\showDOI{\tempurl}


\bibitem[Merckx(2024)]%
        {juliaMLIRFrontend}
\bibfield{author}{\bibinfo{person}{Jules Merckx}.}
  \bibinfo{year}{2024}\natexlab{}.
\newblock \bibinfo{title}{Building Bridges: Julia as an MLIR Frontend}.
\newblock
\newblock


\bibitem[Moses and Churavy(2020)]%
        {enzyme}
\bibfield{author}{\bibinfo{person}{William Moses} {and}
  \bibinfo{person}{Valentin Churavy}.} \bibinfo{year}{2020}\natexlab{}.
\newblock \showarticletitle{Instead of Rewriting Foreign Code for Machine
  Learning, Automatically Synthesize Fast Gradients}. In
  \bibinfo{booktitle}{\emph{Advances in Neural Information Processing
  Systems}}, \bibfield{editor}{\bibinfo{person}{H.~Larochelle},
  \bibinfo{person}{M.~Ranzato}, \bibinfo{person}{R.~Hadsell},
  \bibinfo{person}{M.~F. Balcan}, {and} \bibinfo{person}{H.~Lin}} (Eds.),
  Vol.~\bibinfo{volume}{33}. \bibinfo{publisher}{Curran Associates, Inc.},
  \bibinfo{pages}{12472--12485}.
\newblock
\urldef\tempurl%
\url{https://proceedings.neurips.cc/paper/2020/file/9332c513ef44b682e9347822c2e457ac-Paper.pdf}
\showURL{%
\tempurl}


\bibitem[Qin et~al\mbox{.}(2024)]%
        {DBLP:journals/pacmpl/QinOGHKS24}
\bibfield{author}{\bibinfo{person}{Xueying Qin}, \bibinfo{person}{Liam
  O'Connor}, \bibinfo{person}{Rob van Glabbeek}, \bibinfo{person}{Peter
  H{\"{o}}fner}, \bibinfo{person}{Ohad Kammar}, {and} \bibinfo{person}{Michel
  Steuwer}.} \bibinfo{year}{2024}\natexlab{}.
\newblock \showarticletitle{Shoggoth: {A} Formal Foundation for Strategic
  Rewriting}.
\newblock \bibinfo{journal}{\emph{Proc. {ACM} Program. Lang.}}
  \bibinfo{volume}{8}, \bibinfo{number}{{POPL}} (\bibinfo{year}{2024}),
  \bibinfo{pages}{61--89}.
\newblock
\urldef\tempurl%
\url{https://doi.org/10.1145/3633211}
\showDOI{\tempurl}


\bibitem[Radford et~al\mbox{.}(2023)]%
        {pmlr-v202-radford23a}
\bibfield{author}{\bibinfo{person}{Alec Radford}, \bibinfo{person}{Jong~Wook
  Kim}, \bibinfo{person}{Tao Xu}, \bibinfo{person}{Greg Brockman},
  \bibinfo{person}{Christine Mcleavey}, {and} \bibinfo{person}{Ilya
  Sutskever}.} \bibinfo{year}{2023}\natexlab{}.
\newblock \showarticletitle{Robust Speech Recognition via Large-Scale Weak
  Supervision}. In \bibinfo{booktitle}{\emph{Proceedings of the 40th
  International Conference on Machine Learning}}
  \emph{(\bibinfo{series}{Proceedings of Machine Learning Research},
  Vol.~\bibinfo{volume}{202})}, \bibfield{editor}{\bibinfo{person}{Andreas
  Krause}, \bibinfo{person}{Emma Brunskill}, \bibinfo{person}{Kyunghyun Cho},
  \bibinfo{person}{Barbara Engelhardt}, \bibinfo{person}{Sivan Sabato}, {and}
  \bibinfo{person}{Jonathan Scarlett}} (Eds.). \bibinfo{publisher}{PMLR},
  \bibinfo{pages}{28492--28518}.
\newblock
\urldef\tempurl%
\url{https://proceedings.mlr.press/v202/radford23a.html}
\showURL{%
\tempurl}


\bibitem[Radford et~al\mbox{.}(2019)]%
        {radford2019language}
\bibfield{author}{\bibinfo{person}{Alec Radford}, \bibinfo{person}{Jeff Wu},
  \bibinfo{person}{Rewon Child}, \bibinfo{person}{David Luan},
  \bibinfo{person}{Dario Amodei}, {and} \bibinfo{person}{Ilya Sutskever}.}
  \bibinfo{year}{2019}\natexlab{}.
\newblock \showarticletitle{Language Models are Unsupervised Multitask
  Learners}.
\newblock  (\bibinfo{year}{2019}).
\newblock


\bibitem[Ragan-Kelley et~al\mbox{.}(2013)]%
        {halide}
\bibfield{author}{\bibinfo{person}{Jonathan Ragan-Kelley},
  \bibinfo{person}{Connelly Barnes}, \bibinfo{person}{Andrew Adams},
  \bibinfo{person}{Sylvain Paris}, \bibinfo{person}{Fr{\'e}do Durand}, {and}
  \bibinfo{person}{Saman Amarasinghe}.} \bibinfo{year}{2013}\natexlab{}.
\newblock \showarticletitle{Halide: a language and compiler for optimizing
  parallelism, locality, and recomputation in image processing pipelines}.
\newblock \bibinfo{journal}{\emph{Acm Sigplan Notices}} \bibinfo{volume}{48},
  \bibinfo{number}{6} (\bibinfo{year}{2013}), \bibinfo{pages}{519--530}.
\newblock


\bibitem[Seeker et~al\mbox{.}(2024)]%
        {DBLP:conf/cgo/SeekerCCFHL24}
\bibfield{author}{\bibinfo{person}{Volker Seeker}, \bibinfo{person}{Chris
  Cummins}, \bibinfo{person}{Murray Cole}, \bibinfo{person}{Bj{\"{o}}rn
  Franke}, \bibinfo{person}{Kim~M. Hazelwood}, {and} \bibinfo{person}{Hugh
  Leather}.} \bibinfo{year}{2024}\natexlab{}.
\newblock \showarticletitle{Revealing Compiler Heuristics Through Automated
  Discovery and Optimization}. In \bibinfo{booktitle}{\emph{{IEEE/ACM}
  International Symposium on Code Generation and Optimization, {CGO} 2024,
  Edinburgh, United Kingdom, March 2-6, 2024}},
  \bibfield{editor}{\bibinfo{person}{Tobias Grosser},
  \bibinfo{person}{Christophe Dubach}, \bibinfo{person}{Michel Steuwer},
  \bibinfo{person}{Jingling Xue}, \bibinfo{person}{Guilherme Ottoni}, {and}
  \bibinfo{person}{ernando Magno Quint{\~{a}}o~Pereira}} (Eds.).
  \bibinfo{publisher}{{IEEE}}, \bibinfo{pages}{55--66}.
\newblock
\urldef\tempurl%
\url{https://doi.org/10.1109/CGO57630.2024.10444847}
\showDOI{\tempurl}


\bibitem[Steuwer et~al\mbox{.}(2017)]%
        {lift}
\bibfield{author}{\bibinfo{person}{Michel Steuwer}, \bibinfo{person}{Toomas
  Remmelg}, {and} \bibinfo{person}{Christophe Dubach}.}
  \bibinfo{year}{2017}\natexlab{}.
\newblock \showarticletitle{Lift: a functional data-parallel {IR} for
  high-performance {GPU} code generation}. In
  \bibinfo{booktitle}{\emph{Proceedings of the 2017 International Symposium on
  Code Generation and Optimization, {CGO} 2017, Austin, TX, USA, February 4-8,
  2017}}, \bibfield{editor}{\bibinfo{person}{Vijay~Janapa Reddi},
  \bibinfo{person}{Aaron Smith}, {and} \bibinfo{person}{Lingjia Tang}} (Eds.).
  \bibinfo{publisher}{{ACM}}, \bibinfo{pages}{74--85}.
\newblock
\urldef\tempurl%
\url{http://dl.acm.org/citation.cfm?id=3049841}
\showURL{%
\tempurl}


\bibitem[Sun et~al\mbox{.}(2020)]%
        {sun2020mobilebert}
\bibfield{author}{\bibinfo{person}{Zhiqing Sun}, \bibinfo{person}{Hongkun Yu},
  \bibinfo{person}{Xiaodan Song}, \bibinfo{person}{Renjie Liu},
  \bibinfo{person}{Yiming Yang}, {and} \bibinfo{person}{Denny Zhou}.}
  \bibinfo{year}{2020}\natexlab{}.
\newblock \bibinfo{title}{MobileBERT: a Compact Task-Agnostic BERT for
  Resource-Limited Devices}.
\newblock
\newblock
\showeprint[arxiv]{2004.02984}~[cs.CL]


\bibitem[Vasilache et~al\mbox{.}(2022)]%
        {linalg}
\bibfield{author}{\bibinfo{person}{Nicolas Vasilache},
  \bibinfo{person}{Oleksandr Zinenko}, \bibinfo{person}{Aart~JC Bik},
  \bibinfo{person}{Mahesh Ravishankar}, \bibinfo{person}{Thomas Raoux},
  \bibinfo{person}{Alexander Belyaev}, \bibinfo{person}{Matthias Springer},
  \bibinfo{person}{Tobias Gysi}, \bibinfo{person}{Diego Caballero},
  \bibinfo{person}{Stephan Herhut}, {et~al\mbox{.}}}
  \bibinfo{year}{2022}\natexlab{}.
\newblock \showarticletitle{Structured Operations: Modular Design of Code
  Generators for Tensor Compilers}. In \bibinfo{booktitle}{\emph{International
  Workshop on Languages and Compilers for Parallel Computing}}. Springer,
  \bibinfo{pages}{141--156}.
\newblock


\bibitem[Willsey et~al\mbox{.}(2021)]%
        {egg}
\bibfield{author}{\bibinfo{person}{Max Willsey}, \bibinfo{person}{Chandrakana
  Nandi}, \bibinfo{person}{Yisu~Remy Wang}, \bibinfo{person}{Oliver Flatt},
  \bibinfo{person}{Zachary Tatlock}, {and} \bibinfo{person}{Pavel Panchekha}.}
  \bibinfo{year}{2021}\natexlab{}.
\newblock \showarticletitle{egg: Fast and extensible equality saturation}.
\newblock \bibinfo{journal}{\emph{Proc. {ACM} Program. Lang.}}
  \bibinfo{volume}{5}, \bibinfo{number}{POPL}, Article \bibinfo{articleno}{23}
  (\bibinfo{date}{jan} \bibinfo{year}{2021}), \bibinfo{numpages}{29}~pages.
\newblock
\urldef\tempurl%
\url{https://doi.org/10.1145/3434304}
\showDOI{\tempurl}


\bibitem[Zheng et~al\mbox{.}(2020)]%
        {zheng2020ansor}
\bibfield{author}{\bibinfo{person}{Lianmin Zheng}, \bibinfo{person}{Chengfan
  Jia}, \bibinfo{person}{Minmin Sun}, \bibinfo{person}{Zhao Wu},
  \bibinfo{person}{Cody~Hao Yu}, \bibinfo{person}{Ameer Haj-Ali},
  \bibinfo{person}{Yida Wang}, \bibinfo{person}{Jun Yang},
  \bibinfo{person}{Danyang Zhuo}, \bibinfo{person}{Koushik Sen},
  {et~al\mbox{.}}} \bibinfo{year}{2020}\natexlab{}.
\newblock \showarticletitle{Ansor: Generating $\{$High-Performance$\}$ tensor
  programs for deep learning}. In \bibinfo{booktitle}{\emph{14th USENIX
  symposium on operating systems design and implementation (OSDI 20)}}.
  \bibinfo{pages}{863--879}.
\newblock


\bibitem[Zinenko(2023)]%
        {mlir-not-ml-compiler}
\bibfield{author}{\bibinfo{person}{Oleksandr Zinenko}.}
  \bibinfo{year}{2023}\natexlab{}.
\newblock \bibinfo{title}{MLIR Is Not an ML Compiler -- And Other Common
  Misconceptions}.
\newblock
\newblock
\urldef\tempurl%
\url{https://llvm.org/devmtg/2023-10/slides/techtalks/Zinenko-MLIRisNotAnMLCompiler.pdf}
\showURL{%
\tempurl}


\end{thebibliography}

\end{document}